\documentclass[conference]{IEEEtran}
\IEEEoverridecommandlockouts{}
\usepackage{cite}
\usepackage{amsmath,amssymb,amsfonts}
\usepackage{algorithmic}
\usepackage{graphicx}
\usepackage{textcomp}
\usepackage{xcolor}
\usepackage{hyperref}
\usepackage{diffcoeff}
\usepackage{booktabs}
\usepackage{siunitx}
\def\BibTeX{{\rm B\kern-.05em{\sc i\kern-.025em b}\kern-.08em
    T\kern-.1667em\lower.7ex\hbox{E}\kern-.125emX}}

\newcommand{\ui}[2]{#1_{\text{#2}}}  % upright sub-index
\begin{document}

\title{Deep Dictionary-Free Method for Identifying Linear Model of Nonlinear System with Input Delay\\

    \thanks{The authors acknowledge the contribution of the Program to support young researchers at STU under the projects AIOperator: eXplainable Intelligence for Industry (AXII) and Koopman Operator in Process Control. The authors gratefully acknowledge the contribution of the Scientific Grant Agency of the Slovak Republic under the grants VEGA 1/0490/23 and 1/0239/24, the Slovak Research and Development Agency under the project APVV--21--0019. This research is funded by the Horizon Europe under the grant no. 101079342 (Fostering Opportunities Towards Slovak Excellence in Advanced Control for Smart Industries).}
}

\author{\IEEEauthorblockN{Patrik Val\'{a}bek, Marek Wadinger, Michal Kvasnica, Martin Klau\v{c}o$^\dagger$}
\IEEEauthorblockA{\textit{Institute of Information Engineering, Automation, and Mathematics} \\
    \textit{Slovak University of Technology in Bratislava}\\
    Bratislava, Slovakia \\
    \texttt{patrik.valabek@stuba.sk}}
\thanks{$\dagger$Martin Klau\v{c}o (\texttt{martin.klauco@cvut.cz}) is also with the Department of Control Engineering, Czech Technical University in Prague. M. Klaučo is supported by the European Union project ROBOPROX (Reg. No. CZ.02.01.01/00/22\_008/0004590). The manuscript originates with Slovak University of Technology in Bratislava, while its final submission was done in cooperation with CTU in Prague. }
}

\maketitle

\begin{abstract}
    %Nonlinear dynamical systems with input delays pose significant challenges for prediction, estimation, and control due to their inherent complexity and the impact of delays on system behavior. Traditional linear control techniques often fail in these contexts, necessitating innovative approaches. This paper presents a dictionary-free method for learning linear representations of nonlinear systems with input delays using deep neural networks (DNNs). Leveraging the Koopman operator framework, which globally linearizes nonlinear dynamics, we address the limitations of extended Dynamic Mode Decomposition (eDMD). While eDMD enhances model capability with nonlinear measurements, its reliance on predefined dictionaries constrains its accuracy. Our approach uses DNNs to automatically generate and update nonlinear transformations, enabling the learning of high-fidelity Koopman operator models. Additionally, we incorporate time-delayed embeddings to account for input delays, ensuring precise modeling and improved long-term forecasting of nonlinear dynamics. Our method provides a robust framework for modeling nonlinear systems with input delays, offering significant advancements over existing techniques.
    Nonlinear dynamical systems with input delays pose significant challenges for prediction, estimation, and control due to their inherent complexity and the impact of delays on system behavior. Traditional linear control techniques often fail in these contexts, necessitating innovative approaches.
    This paper introduces a novel approach to approximate the Koopman operator using an LSTM-enhanced Deep Koopman model, enabling linear representations of nonlinear systems with time delays. By incorporating Long Short-Term Memory (LSTM) layers, the proposed framework captures historical dependencies and efficiently encodes time-delayed system dynamics into a latent space. Unlike traditional extended Dynamic Mode Decomposition (eDMD) approaches that rely on predefined dictionaries, the LSTM-enhanced Deep Koopman model is dictionary-free, which mitigates the problems with the underlying dynamics being known and incorporated into the dictionary.
    Quantitative comparisons with extended eDMD on a simulated system demonstrate highly significant performance gains in prediction accuracy in cases where the true nonlinear dynamics are unknown and achieve comparable results to eDMD with known dynamics of a system.
\end{abstract}


\section{Introduction}
Understanding and controlling nonlinear dynamical systems is a fundamental challenge across various scientific and engineering disciplines. Traditional linear control techniques often fall short when applied to such systems due to their inherent nonlinearity. This becomes even more challenging in time-delayed systems, where the presence of delays introduces additional complexity by coupling past states with the current system dynamics. These delays can significantly degrade control performance and stability if not adequately accounted for. A promising approach to this problem is the identification of coordinate transformations that render nonlinear dynamics approximately linear, thus enabling the application of linear theory for prediction, estimation, and control.

Koopman analysis, a technique that facilitates the linearization of nonlinear systems through the Koopman operator, has gained considerable traction in recent years~\cite{Wilson2023siamkoopman, Mauroy2016cdc}. The Koopman operator offers a global linearization of dynamics by mapping the original nonlinear system into a higher-dimensional space where the dynamics can be represented linearly~\cite{Mezic2004101, Mezić2005}.
This method is particularly appealing because it shifts the complexity from the system's nonlinear equations to the eigenfunctions of the Koopman operator. These eigenfunctions span an invariant subspace, enabling the system's dynamics to be represented by a finite-dimensional matrix within this subspace, simplifying the analysis and computation of the dynamics.

Finite-dimensional approximations of the Koopman operator are often achieved using Dynamic Mode Decomposition (DMD)~\cite{schmid2010dynamic}. While DMD identifies spatio-temporal coherent structures from high-dimensional systems, it typically fails to capture nonlinear transients due to its reliance on linear measurements. To address this, Extended DMD (eDMD)~\cite{williams2015data} incorporates a dictionary, improving its capability to model nonlinear systems. In this context dictionary stands for a collection or set of basis functions or candidate models used to represent or approximate the nonlinear relationships. However, eDMD faces challenges such as high dimensionality and closure issues, which arise because there is no guarantee that the nonlinear measurements form a Koopman invariant subspace~\cite{Lusch2018}. Consequently, the identification and representation of Koopman eigenfunctions remain crucial tasks, motivating the use of advanced deep learning techniques.

To overcome the limitations of eDMD, deep neural networks (DNNs) have been employed to learn Koopman operator representations. DNNs remove the bottleneck of predefined dictionaries by creating linear representations of a system through nonlinear transformations of individual neurons. These neurons combine to form complex functions parametrized by tunable weights and biases, allowing the network to adaptively learn the optimal transformations during training. This approach significantly enhances the fidelity of Koopman operator models, particularly in multi-step prediction tasks, thus improving long-term forecasting of nonlinear dynamics~\cite{Yeung2019}.
\begin{figure*}
    \centerline{\includegraphics[width = \linewidth]{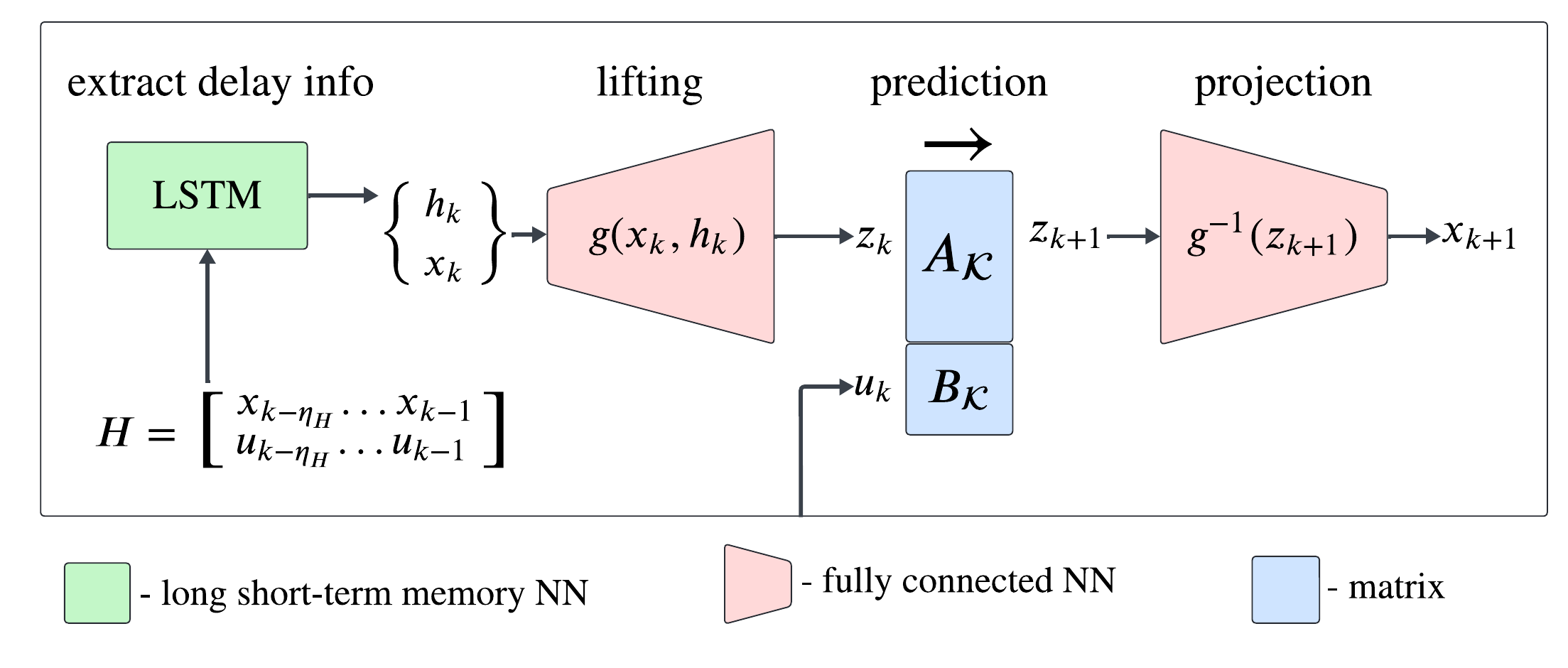}}
    \caption{The LSTM-enhanced Deep Koopman algorithm architecture. The model is built on the autoencoder architecture of the Deep Koopman Operator. The history of state and input data is first extracted using the LSTM layer. The hidden states of the last LSTM layer are then concatenated with the last state as input to the autoencoder part. After lifting the states, there is a prediction layer represented with the matrices \(A_\mathcal{K}\) and \(B_\mathcal{K}\). After the prediction layer, the lifted states are projected back to states using the decoder part of the autoencoder.}
    \label{fig:introduction:LSTM-enhanced_deep_koopman_scheme}
\end{figure*}

An essential aspect of modeling dynamical systems is accounting for time delays, which are common in many physical and industrial processes. Time delay can significantly affect system behavior and control performance. Introducing time-delayed embeddings of control action in DMD improves the identification of the system with input delays while not explicitly identifying the time delays. Various methods exist to identify and model time delays, including state-space realization approaches and correlation analysis. The state-space realization approach involves constructing a state-space representation based on input-output data from experiments, extracting time-delay information from the system's impulse response~\cite{Lima2015254}. Correlation analysis, on the other hand, focuses on identifying time delays in dynamic processes with disturbances, aiming to improve control accuracy and stability~\cite{Li201792}. The limitations of correlation analysis addressed by eDMD with time-delayed embeddings include its inability to account for system dynamics, as eDMD captures the full state evolution through time-delay coordinates, and its sensitivity to disturbances and noise, as eDMD leverages a higher-dimensional representation to better isolate underlying dynamics.

This paper presents a dictionary-free method for learning linear representations of nonlinear systems with input delays using deep neural networks. By leveraging the strengths of deep learning in generating and updating nonlinear transformations, our approach aims to overcome the limitations of traditional Koopman operator methods and provide a robust framework for modeling and controlling nonlinear systems with time delays. The following sections are structured to guide the reader through the theoretical foundations, methodological advancements, and practical insights of our approach, addressing key challenges and presenting proposed solutions in this domain.

In Section~\ref{sec:methodology}, we detail the methodological framework of our approach. Subsection~\ref{subsec:methodology:lstm_koopman} introduces the implementation of the LSTM-enhanced Deep Koopman model, emphasizing its ability to capture nonlinear dynamics through latent space representations, where LSTM refers to deep neural network architecture Long-Short Term Memory, which is a type of recurrent neural network. Subsection~\ref{subsec:methodology:explicit_loss} then describes the explicit loss function design, focusing on how it ensures accurate prediction and stability in the learning process.

Section~\ref{sec:results} presents the results of our approach, providing quantitative and qualitative evaluations. Subsection~\ref{subsec:results:comparison_edmd} compares our method with the established extended Dynamic Mode Decomposition (eDMD), highlighting improvements in capturing time-delayed nonlinear dynamics and demonstrating the advantages of our approach in representing unknown dynamics.

\section{LSTM-Enhanced Deep Koopman Approach}\label{sec:methodology}
% alebo Data-Driven Koopman Operator Approximation

The foundation of the proposed approach lies in approximating the Koopman operator, a linear operator that represents the dynamics of nonlinear systems in an embedded space.
Formally, we aim to find the underlying linear dynamics in state space equation covering
\begin{equation}
    z_{k+1} = A_\mathcal{K}z_k + B_\mathcal{K}u_k.
\end{equation}

where the Koopman matrices \(A_\mathcal{K} \) and \( B_\mathcal{K} \) are the solution for the optimization problem:

\begin{equation}
    \mathrm{argmin}_{\mathcal{K}} \sum_{k=0}^{m} \| z_{k+1} - A_\mathcal{K}z_{k} - B_\mathcal{K}u_k \|_F,
\end{equation}

where \( z_{k} \in \mathbb{R}^n\) represent the vector of lifted states and \( z_{k+1} \) corresponding states in next step, respectively, \( \|\cdot \| _F\) denotes the Frobenius norm and \(m\) denotes the number of data snapshots. This formulation provides the basis for deriving a linear approximation of the underlying dynamics, even for systems with time delays.

The proposed approach leveraging the power of a long short-term memory (LSTM) layer encoding the history of a system is a novel approach for data-based identification of approximation of the Koopman operator. It can accurately identify the linear model for processes with time delays. Its basis comes from the Deep Koopman Operator (DKO)~\cite{lusch2018deep} approach.

LSTM-enhanced Deep Koopman is built to approximate the Koopman operator for the systems with time delays. Thanks to the LSTM layer that processes the system's history denoted \(H\), we can effectively capture the time delays in the system. Time delays are encoded as the hidden states of the last LSTM layer denoted \(h_k\). Thanks to this encoding (hidden states of LSTM), we can obtain the relevant information about the system's past.

For example, using the LSTM layer, we can obtain the chosen number (this is a hyperparameter) of hidden states. These hidden states are then used as input to the encoder part of our approach. This way, we can reduce the amount of input data to the model compared to the Deep Koopman model, which would take the whole history of the system as input. This would be very inefficient and computationally more expensive.

Also, compared to the DMD, where we would use the history of measurements, we can have a smaller Koopman matrix \(\mathcal{K}\). Considering, for example, the last \(\ui{l}{his}\) measurements of the system states \(x \in \mathbb{R}^n\), the \(A_\mathcal{K}\) would have the size \(\ui{l}{his}n \times \ui{l}{his}n\) and the \(B_\mathcal{K}\) would have the size \(\ui{l}{his}n \times m\),  where \(m\) is the number of inputs to the system \(u \in \mathbb{R}^m\). This is often a matrix that is more expensive to compute and store if we are dealing with a long history and many states compared to the LSTM-enhanced deep Koopman approach. On the other hand, the LSTM layer can capture the relevant information about the system's past and reduce the amount of input data in the model.

\subsection{Implementation of LSTM-enhanced Deep Koopman}\label{subsec:methodology:lstm_koopman}
LSTM-enhanced Deep Koopman was implemented in the Python programming language using the NeuroMANCER~\cite{Neuromancer2023} library.
%NeuroMANCER stands for Neural Modules with Adaptive Nonlinear Constraints and Efficient Regularizations. It is an open-source differentiable programming (DP) library for solving parametric constrained optimization problems, physics-informed system identification, and parametric model-based optimal control. NeuroMANCER is written in PyTorch and allows for systematically integrating machine learning with scientific computing to create end-to-end differentiable models and algorithms embedded with prior knowledge and physics.
The NeuroMANCER library does not provide the use of LSTM layers, so we implemented them into the library.

%There exists ``Learning Stable Deep Koopman Operators'' and ``Learning Stable Control-oriented Deep Koopman Operators'' examples based on~\cite{shi2022deep, korda2020optimal,lusch2018deep} which served as a base for our implementation of the LSTM-enhanced Deep Koopman model.
%Also, they provided implementation for learning stable Koopman operator based on Generic Stable Layers~\cite{skomski2021constrained, drgovna2022dissipative, zhang2018stabilizing} for constrained learning of NN\@. Also, implementing the control input in the model needed to be changed. Both models were implemented with the control input not encoded to be further compatible with Koopman MPC~\cite{korda2018linear}.

The schematic of LSTM-enhanced Deep Koopman architecture can be seen in Fig.~\ref{fig:introduction:LSTM-enhanced_deep_koopman_scheme}. The model is built on the autoencoder architecture of the DKO\@. Before the prediction we compute the history \(H\) of our system:
\begin{equation}
    H = \begin{bmatrix}
        x_{k - \eta_H} & \cdots & x_{k-1} \\
        u_{k - \eta_H} & \cdots & u_{k-1}
    \end{bmatrix},
\end{equation}
where \(\eta_H\) is the number of time steps in the history of the system. We set this parameter to be higher value as our observed or estimated time delay.
The \(H\), consisting of state and input data, is first encoded using the LSTM layer. The hidden states of the last LSTM layer are then concatenated with the current state as input to the autoencoder part. The autoencoder part remains the same as for the DKO\@.

\subsection{Explicit Loss Function}\label{subsec:methodology:explicit_loss}
In the training of the LSTM-enhanced Deep Koopman model, we are also using the same loss function as in the training of the DKO model.
This section is important for the reproducibility of the results.
The loss function combines the reconstruction loss, one-step output prediction loss, output trajectory prediction loss, and latent trajectory prediction loss. These loss functions are described in the~\cite{lusch2018deep}. This function is one of the most important parts of the training process. Compared to DMD, where the loss function is only the mean squared error between the predicted and true future states for a one-step prediction, the Deep Koopman methods also use a multi-step prediction in a loss function. In identifying a system, by using the multi-step prediction, we can better capture the system's dynamics with input delays as opposed to only one-step prediction loss.

\paragraph*{Reconstruction loss}
is the mean squared error between the state \(x_k\) and the encoded (lifted) and decoded (unlifted) state \(g^{-1}(g(x_k,h_k))\):
\begin{equation}
    \mathcal{L}_{\text{rec}} = \left \|x_k - g^{-1}(g(x_k,h_k))\right \|^2.
\end{equation}
This loss is the basic loss function for the autoencoder part of the model. It ensures that we can correctly encode and decode the state of the system.

\paragraph*{One step output prediction loss}
is the mean squared error between the predicted state \(\hat{x}_{k+1}\) and the true state \(x_{k+1}\):

\begin{subequations}
    \begin{align}
        \mathcal{L}_{\text{step}} & = \left \|\hat{x}_{k+1} - x_{k+1}\right \|^2,         \\
        \hat{x}_{k+1}             & = g^{-1}(A_\mathcal{K}g(x_k,h_k) + B_\mathcal{K}u_k),
    \end{align}
\end{subequations}
where \(u_k\) is the input flow rate at time \(k\), the \(A_\mathcal{K}\) and \(B_\mathcal{K}\) are the identified Koopman matrices.

\paragraph*{Output trajectory prediction loss}
\label{par:methodology:output_loss}
is the mean squared error between the predicted trajectory \(\hat{x}_{k+1:k+\ui{N}{L}}\) and the true trajectory \(x_{k+1:k+\ui{N}{L}}\):
\begin{subequations}
    \begin{align}
        \mathcal{L}_{\text{pred}} & = \sum_{i=1}^{\ui{N}{L}}\left \|\hat{x}_{k+i} - x_{k+i}\right \|^2, \\
        \hat{x}_{k+i}             & = g^{-1}\left(\hat{z}_{k+i}\right),                                 \\
        \hat{z}_{k+i}             & = A_{\mathcal{K}}\hat{z}_{k+i-1}+B_{\mathcal{K}}u_{k+i-1},          \\
        \hat{z}_{k}               & = g(x_{k},h_k),
    \end{align}
\end{subequations}
where \(\ui{N}{L}\) is the prediction horizon in loss function specified by a user. In this case, the predicted trajectory \(\hat{x}_{k:k+\ui{N}{L}}\) is predicted based on the \(x_k,~h_k\) and the input trajectory \(u_{k:k+\ui{N}{L}-1}\). This loss function is used to capture the dynamics of the system and also input delays. By using the multi-step prediction loss, we can better capture the dynamics of the system as opposed to only one-step prediction loss.

\paragraph*{Latent trajectory prediction loss}
is the mean squared error between the lifted predicted trajectory \(\hat{z}_{k+1:k+\ui{N}{L}}\) and the lifted true trajectory \(z_{k+1:k+\ui{N}{L}}\):
\begin{subequations}
    \begin{align}
        \mathcal{L}_{\text{lpred}} & = \sum_{k=1}^{\ui{N}{L}}\left \|\hat{z}_{k+i} - z_{k+i}\right \|^2, \\
        z_{k+i}                    & = g(x_{k+i},h_{k+i}),                                               \\
        \hat{z}_{k+i}              & = A_{\mathcal{K}}\hat{z}_{k+i-1}+B_{\mathcal{K}}u_{k+i-1},          \\
        \hat{z}_{k}                & = g(x_{k},h_k),
    \end{align}
\end{subequations}
where this loss function works with the same principle as the output trajectory prediction loss~\ref{par:methodology:output_loss}, but in the space of lifted states.

The final loss function is a weighted sum of all the loss functions:
\begin{equation}
    \label{eq:loss_function}
    \mathcal{L} = \ui{w}{rec}\mathcal{L}_{\text{rec}} + \ui{w}{step}\mathcal{L}_{\text{step}} + \ui{w}{pred}\mathcal{L}_{\text{pred}} + \ui{w}{lpred}\mathcal{L}_{\text{lpred}},
\end{equation}
where \(\ui{w}{rec}\), \(\ui{w}{step}\), \(\ui{w}{pred}\), and \(\ui{w}{lpred}\) are the weights for the reconstruction loss, one-step output prediction loss, output trajectory prediction loss, and latent trajectory prediction loss, respectively. They are set by the user and are used to balance the importance of the loss functions in the training process.

\section{Case Study}\label{sec:results}

\subsection{Comparison with eDMD}\label{subsec:results:comparison_edmd}

This section presentes the comparison of the performance of LSTM-enhanced Deep Koopman with eDMD on a simulated two-tank system without interaction with input delays. The system is described by the following equations:

\begin{equation}
    \begin{aligned}
        \diff{h_1(t)}{t} & = q(t-\tau) - \frac{k_1}{F_1}\sqrt{h_1(t)}                     \\
        \diff{h_2(t)}{t} & = \frac{k_1}{F_2}\sqrt{h_1(t)} - \frac{k_2}{F_2}\sqrt{h_2(t)},
    \end{aligned}\label{eq:two_tanks_system}
\end{equation}

where \(h_1(t)\) and \(h_2(t)\) are the water levels in tanks 1 and 2, respectively, \(q(t)\) is the input flow rate, \(\tau \) is the input delay, and \(k_1\), \(k_2\) are the flow rate constants of the valves and \(F_1\), \(F_2\) are the cross-sectional areas of the tanks.

The experimental data were acquired by simulating this system at sampling period \( \ui{T}{s} = 10 \, \mathrm{s} \) with input delay \( \tau = 20\ui{T}{s} \). Random change in the input flow rate from the interval \( \ui{q}{\text{min}} = 0.0 \, \unit{m^3s^{-1}} \), \( \ui{q}{\text{max}} = 0.03 \) \( \unit{m^3s^{-1}} \) was applied to the system. The simulation yield \( 4 \cdot 10^5\) samples. Gaussian noise with standard deviation of 0.1 was added to the data to simulate real-world conditions.

The data were split into training and testing sets with a ratio of 50:50. The training set was used to train the models, while the testing set was used to evaluate their performance. The models were trained to predict the water levels in tanks \(1\) and \(2\) for the whole testing set based on the applied input flow rate and the initial state of the system.

For reference, we use eDMD with dictionary of polynomials up to degree 2, the correct governing nonlinear dynamics, which is square root of the water levels and the time-delayed embeddings of the input flow rate composed of the previous 20 samples. For comparison, eDMD without the square root of the water levels was also used, simulating a situation where the true nonlinear terms are not known. Therefore, they are not included in the dictionary, which is often the case in more complex systems. Lifted states were obtained by concatenating the dictionary terms and time-delayed embeddings, and the Koopman matrix was learned using eDMD\@.

The LSTM-enhanced Deep Koopman model consists of an LSTM layer with one hidden layer with \(8\) units to extract information about the time delays in the system from the history of the system. The concatenated information is then transformed into lifted states using an encoder layer with a fully connected network. The lifting network consists of input layer with \(10\) neurons representing inputs \(x\text{ and } h\), two hidden layers, each with \(60\) neurons and lifted state layer consisting of \(40\) neurons representing lifted states \(z\).  After lifting the states, there is a prediction layer represented with the matrices \(A_\mathcal{K}\) \(\left(40 \times 40\right)\) and \(B_\mathcal{K}\) \(\left(40 \times 1\right)\). After the prediction layer, the lifted states are projected back to \(2\) states representing the water level in the tanks, using the decoder part of the autoencoder with two hidden layers, each containing \(60\) neurons.
For the weights in \eqref{eq:loss_function} we set \(\ui{w}{rec} = 0\), \(\ui{w}{step} = 1\), \(\ui{w}{pred} = 10\), and \(\ui{w}{lpred} = 1\). The model was trained using PyTorch for \(1500\) epochs with a batch size of \(100\) using the Adam optimizer with a learning rate of \(0.001\).

\begin{figure}[htbp]
    \centerline{\includegraphics[width=\linewidth]{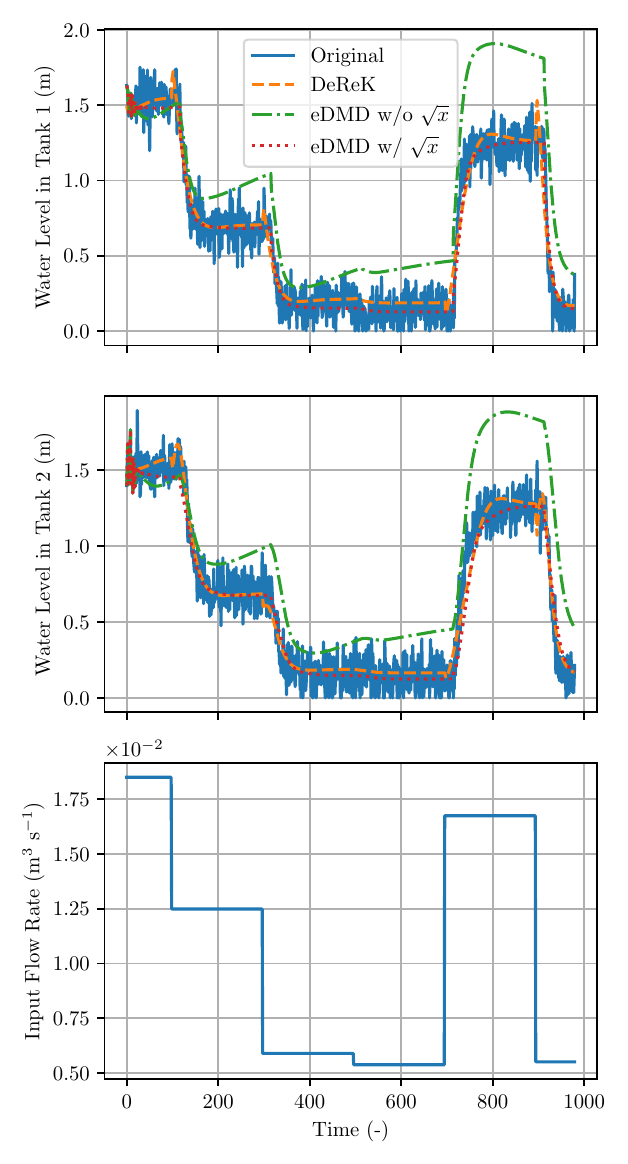}}
    \caption{Predicted and actual water levels in tanks \(1\) and \(2\) for the two tank system. Blue line shows the simulation of~\eqref{eq:two_tanks_system}, while gray shows the noisy data, as used for training. The sequence shows a snippet of testing set which was unseen during system identification.}
    \label{fig:results:two_tanks_results}
\end{figure}

Figure~\ref{fig:results:two_tanks_results} shows the predicted and actual water levels in tanks \(1\) and \(2\) for the two tank systems using LSTM-enhanced Deep Koopman and eDMD with and without the square root of the water levels. All three algorithms were initialized with access to same \(H\). The results demonstrate that LSTM-enhanced Deep Koopman outperforms eDMD, which does not include true nonlinear behavior in terms of prediction accuracy, capturing the system's dynamics more effectively over the testing set.\ eDMD without the square root of the water levels exhibits systematic overestimation of the influence of input flow rate step change, leading to a positive bias in the prediction of water levels. At the same time, LSTM-enhanced Deep Koopman provides more accurate and consistent predictions. The linear model generated by LSTM-enhanced Deep Koopman displays non-minimum phase behavior, which is actually a possible approach in the identification of systems with input delays. Meanwhile, eDMD with the square root of the water levels provides accurate prediction, as it includes the true nonlinear terms in the dictionary, thanks to prior knowledge of the system's dynamics.

The performance of the models was evaluated using the mean absolute error (MAE) between the predicted and actual water levels in tanks 1 and 2. The results are presented in Table~\ref{tab:results:two_tanks_results}. LSTM-enhanced Deep Koopman achieved a significantly lower MAE compared to eDMD without square root, indicating its superior performance in modeling the system with input delays in cases where true nonlinear dynamics are not known. Meanwhile, LSTM-enhanced Deep Koopman provides comparable results to eDMD with square root, demonstrating its effectiveness in capturing the system's dynamics without prior knowledge of the true nonlinear terms, demonstrating the capabilities of the LSTM-enhanced Deep Koopman model in modeling nonlinear systems with input delays. The degradation in performance was, in this case, \( 70\% \) in MAE as compared to eDMD with square root, nevertheless it makes only 0.04 m higher error in absolute water levels. It is important to note that the LSTM-enhanced Deep Koopman model is composed of a smaller number of lifted states compared to eDMD\@. This is thanks to the fact that the history of the system is encoded and does not need to be introduced whole to the model during lifted states generation. This results in an almost \(4\) times smaller Koopman matrix, which is computationally less expensive to compute and store.

\begin{table}[htbp]
    \caption{Performance comparison of used methods with the best performance normed to \(100\% \)}
    \label{tab:results:two_tanks_results}
    \begin{center}
        \begin{tabular}{c S[table-format=0.2] S[table-format=3.2] }
            \toprule
            \textbf{Model}             & \textbf{MAE}~\([m]\) & \textbf{MAE}~\([\%]\) \\
            \midrule
            eDMD (known dynamics)      & 0.175                 & 100                   \\
            LSTM-enhanced Deep Koopman & 0.185                 & 106                   \\
            eDMD (unknown dynamics)    & 0.606                 & 346                  \\
            \bottomrule
        \end{tabular}
    \end{center}
\end{table}

\begin{figure}[htbp]
    \centerline{\includegraphics[width=\linewidth]{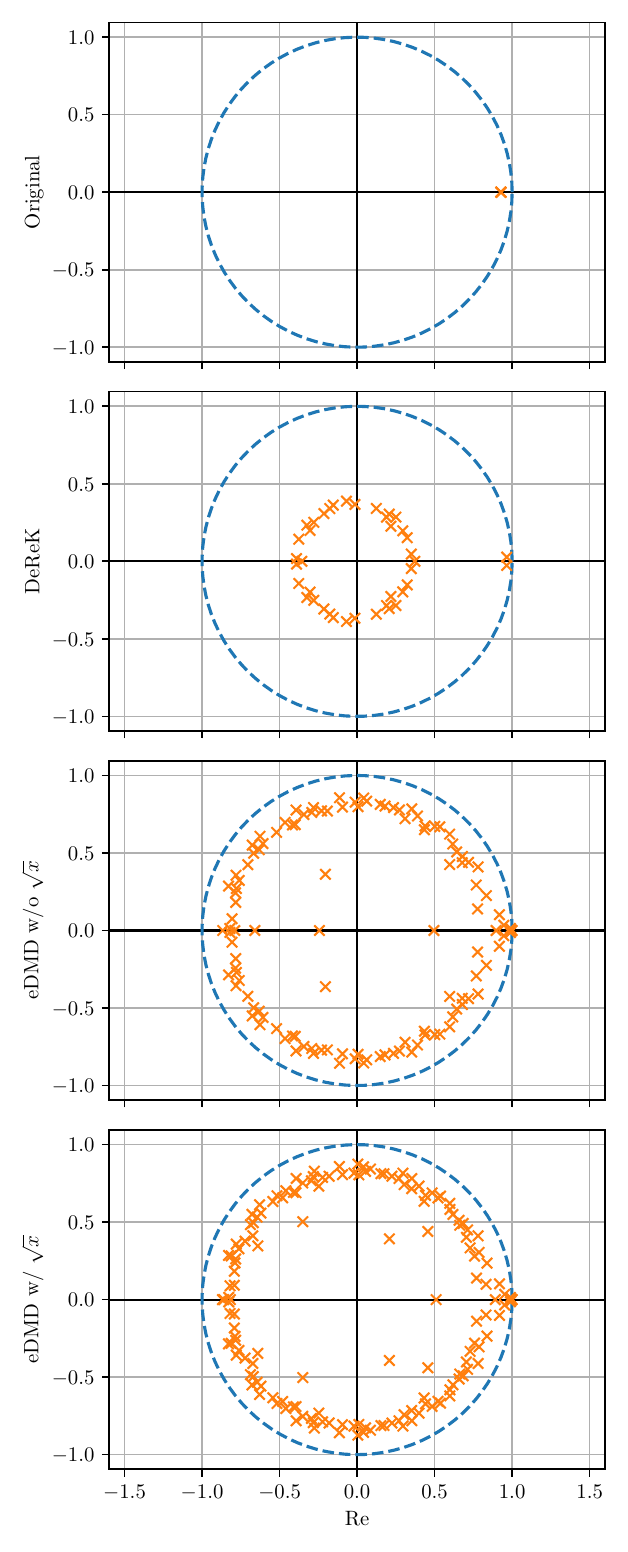}}
    \caption{Comparison of eigenvalues of the Koopman operator for the two tank system. Subplot ``Original'' shows the eigenvalues of the original system, lienarized around the tank levels of 1 m.}
    \label{fig:eigs}
\end{figure}

Lastly, we compare the eigenvalues of the Koopman operator for the two tank systems using LSTM-enhanced Deep Koopman and eDMD\@. Figure~\ref{fig:eigs} shows that LSTM-enhanced Deep Koopman provides more accurate and consistent eigenvalues with original eigenvalues compared to eDMD\@. This is an important contribution as the predictions could better align with reality and provide more accurate and consistent results.

\section{Conclusion}
This paper presents a novel dictionary-free method for identifying linear representations of nonlinear systems with input delays using deep neural networks. The LSTM-enhanced Deep Koopman model leverages the strengths of deep learning to generate and update nonlinear transformations, enabling the learning of high-fidelity Koopman operator models. By incorporating the history of the system, LSTM-enhanced Deep Koopman ensures precise modeling and improved long-term forecasting of nonlinear dynamics with input delays. The results demonstrate that LSTM-enhanced Deep Koopman outperforms traditional eDMD in cases where true dynamics are not anticipated in the dictionary. In this case, the LSTM-enhanced Deep Koopman approach provides improvement more than \(3\) times in MAE, showing the ability to capture the system's dynamics more effectively while delivering accurate and consistent predictions. With eDMD, where true dynamics are known, the LSTM-enhanced Deep Koopman underperforms by only \(6\%\) in MAE, which is a considerable trade considering that the true dynamics are not always known. Future work will focus on extending the LSTM-enhanced Deep Koopman model to more complex systems and exploring its applications in model predictive control.

\bibliographystyle{IEEEtran}
\bibliography{main}

\end{document}